\documentclass[aps, prb, 
reprint, superscriptaddress, nobibnotes, 
amsmath, amssymb, 
]{revtex4-1}

\usepackage{hyperref}
    \hypersetup{
        colorlinks=true, 
        linkcolor=blue,
        urlcolor=blue,
        citecolor=blue,
        bookmarksopen=true,
    }
\usepackage{graphicx}
\usepackage{latexsym}
\usepackage{bm}
\usepackage[alsoload=synchem]{siunitx}

\newcommand*{\Tc}{$T_\text{c}$}
\newcommand*{\Tp}{T$'$}
\newcommand*{\EF}{$E_\text{F}$}

\newcommand*{\wave}{$\sim$}

\newcommand*{\Neel}{N$\acute{\text{e}}$el}
\newcommand*{\hv}{$h\nu$}

\begin{document}

\title{Nature of carrier doping in \texorpdfstring{\Tp{}-${\text{La}_\text{1.8-x}\text{Eu}_\text{0.2}\text{Sr}_\text{x}\text{CuO}_\text{4}}$\\}{LESCO} studied by X-Ray Photoemission and Absorption Spectroscopy}

\author{Chun Lin}
    \thanks{These two authors contributed equally\\Email: clin@wyvern.phys.s.u-tokyo.ac.jp}
\author{Masafumi Horio}
    \thanks{These two authors contributed equally\\Email: clin@wyvern.phys.s.u-tokyo.ac.jp}
    \affiliation{Department of Physics, University of Tokyo, Bunkyo-ku, Tokyo 113-0033, Japan.}
\author{Takayuki Kawamata} 
\author{Shin Saito} 
    \affiliation{Department of Applied Physics, Tohoku University, Sendai 980-8579, Japan.}
\author{Keisuke Koshiishi}
\author{Shoya Sakamoto}
    \affiliation{Department of Physics, University of Tokyo, Bunkyo-ku, Tokyo 113-0033, Japan.}
\author{Yujun Zhang}
\author{Kohei Yamamoto}
\author{Keisuke Ikeda}
\author{Yasuyuki Hirata}
\author{Kou Takubo}
\author{Hiroki Wadati}
    \affiliation{Institute for Solid State Physics, University of Tokyo, Kashiwa, Chiba 277-8581, Japan.}
\author{Akira Yasui}
\author{Yasumasa Takagi}
\author{Eiji Ikenaga}
    \affiliation{Japan Synchrotron Radiation Research Institute, Sayo, Hyogo 679-5198, Japan.}
\author{Tadashi Adachi}
    \affiliation{Department of Engineering and Applied Sciences, Sophia University, Chiyoda, Tokyo 102-8554, Japan.}
\author{Yoji Koike}
    \affiliation{Department of Applied Physics, Tohoku University, Sendai 980-8579, Japan.}
\author{Atsushi Fujimori}
    \affiliation{Department of Physics, University of Tokyo, Bunkyo-ku, Tokyo 113-0033, Japan.}
    \affiliation{Department of Applied Physics, Waseda University, Shinjuku, Tokyo 169-8555, Japan.}


\begin{abstract}
Recently, hole-doped superconducting cuprates with the \Tp{}-structure ${\text{La}_\text{1.8-x}\text{Eu}_\text{0.2}\text{Sr}_\text{x}\text{CuO}_\text{4}}$ (LESCO) have attracted a lot of attention. 
We have performed x-ray photoemission and absorption spectroscopy measurements on as-grown and reduced \Tp{}-LESCO. 
Results show that electrons and holes were doped by reduction annealing and Sr substitution, respectively. 
However, it is shown that the system remains on the electron-doped side of the Mott insulator or that the charge-transfer gap is collapsed in the parent compound.

\end{abstract}

\maketitle

Starting from the so-called 214-type copper oxides with the ${\text{K}_\text{2}\text{Ni}\text{F}_\text{4}}$ type (T-type) and ${\text{Nd}_\text{2}\text{CuO}_\text{4}}$ type (\Tp{}-type) structure, unconventional superconductivity evolves after hole and electron doping, respectively~\cite{Armitage_Rev.Mod.Phys.823_2010}.
In the T-type cuprates, the Cu atom is octahedrally surrounded by six oxygen atoms.
By contrast, the \Tp{}-type cuprates are characterized by the absence of two apical oxygen atoms in the octahedron.
Thanks to the switchable carrier types, the ``214'' cuprate families provides a unique platform to study how carriers are doped into the parent Mott insulators and to understand the long-standing puzzle of the asymmetry in the temperature-doping phase diagram of the cuprates~\cite{Armitage_Rev.Mod.Phys.823_2010}.
However, the different crystal structures between the hole- and electron-doped compounds preclude unambiguous discussion about the similarities and differences between the hole- and electron-doped cuprates.

Recently, superconducting (SC) \Tp{}-type cuprates without Ce doping, such as thin-film ${\text{Nd}_\text{2}\text{CuO}_\text{4}}$~\cite{Matsumoto_PhysicaC:Superconductivity46915_2009} and ${\text{Pr}_\text{2}\text{CuO}_\text{4}}$~\cite{Krockenberger_Sci.Rep.3_2013}, and even with Sr doping such as ${\text{La}_\text{1.8-x}\text{Eu}_\text{0.2}\text{Sr}_\text{x}\text{CuO}_\text{4}}$ (LESCO)~\cite{Takamatsu_Appl.Phys.Express57_2012}, have raised questions about the Mott-insulating nature of the parent compounds.
Nevertheless, in these systems, too, the post-annealing, which is generally believed to reduce the excess apical oxygen~\cite{Armitage_Rev.Mod.Phys.823_2010}, is indispensable to realize the superconductivity in the \Tp{}-type cuprates, indicating possible electron doping through the removal of oxygen atoms.
In fact, the annealing-induced changes in the \Neel{} temperature and optical conductivity in ${\text{Nd}_\text{2-x}\text{Ce}_\text{x}\text{CuO}_\text{4}}$ (NCCO) have been explained by electron doping~\cite{Mang_Phys.Rev.Lett.932_2004,Arima_Phys.Rev.B489_1993}.
Moreover, the estimated electron concentrations in reduced ${\text{Pr}_\text{2-y-x}\text{La}_\text{y}\text{Ce}_\text{x}\text{CuO}_\text{4}}$ were found to be significantly larger than the nominal Ce doping levels~\cite{Horio_Nat.Commun.7_2016}, implying the possibility of electron doping through the removal of oxygen atoms not only from the apical sites (excess oxygens) but also from the regular sites.
Inspired by these studies, hard x-ray photoemission spectroscopy (HAXPES) measurements have been performed on the SC ${\text{Nd}_\text{2}\text{CuO}_\text{4}}$ thin films to detect electron doping through the measurements of the chemical-potential shift deduced from the core-level spectra~\cite{Horio_Phys.Rev.Lett.12025_2018}.
The results, together with subsequent angle-resolved photoemission spectroscopy measurements on SC ${\text{Pr}_\text{2}\text{CuO}_\text{4}}$ thin films to probe the Fermi surface area~\cite{Horio_Phys.Rev.B982_2018}, have unveiled that electrons are actually doped in Ce-undoped \Tp{}-type cuprates after reduction annealing.
On top of the superconductivity without Ce doping, LESCO ($x > 0$) is the first nominally hole-doped SC cuprate with the \Tp{}-type structure~\cite{Takamatsu_Appl.Phys.Express57_2012,Takamatsu_PhysicsProcedia58SupplementC_2014}.
If carrier doping can be controlled both by annealing and Sr substitution, one might be able to dope the same system either with electrons or holes and thereby to observe a transition between the two types of doping across a truly undoped state in-between.

In the present Short Note, we report HAXPES and x-ray absorption spectroscopy (XAS) studies on as-grown and annealed LESCO with $x$ = 0, 0.05, and 0.10.
The results show that electrons and holes were doped by reduction annealing and Sr substitution, respectively, 
while no appreciable chemical-potential jump~\cite{Ikeda_Phys.Rev.B822_2010} was observed in any doping range.
In the XAS spectra, no Zhang-Rice singlet (ZRS) feature~\cite{Pellegrin_Phys.Rev.B476_1993} was found.
Assuming the presence of a charge-transfer (CT) gap~\cite{Ikeda_Phys.Rev.B822_2010}, the current data indicate that there are no doped holes irrespective of Sr content and that the system remains on the electron-doped side of the Mott insulator.
Alternatively, the experimental data can be consistently interpreted within the scenario that the CT gap collapsed~\cite{Adachi_J.Phys.Soc.Jpn.826_2013}. 

\begin{figure}[b]
    \includegraphics[width=0.5\textwidth]{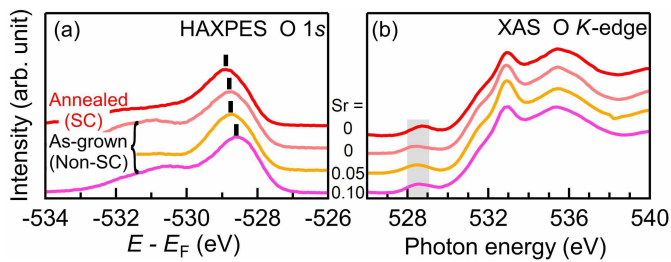}
    \caption{
    O 1\textit{s} hard x-ray photoemission spectroscopy (HAXPES) spectra (a) and O \textit{K}-edge x-ray absorption spectroscopy (XAS) spectra (b) of \Tp{}-${\text{La}_\text{1.8-x}\text{Eu}_\text{0.2}\text{Sr}_\text{x}\text{CuO}_\text{4}}$ with Sr content $x$ = 0 (annealed and as-grown), 0.05 (as-grown), and 0.10 (as-grown). 
    Black bars in (a) indicate the peak positions of each spectrum and the shaded area in (b) highlights the pre-edge peaks at 528 - 529 eV. 
    XAS spectra have been normalized to the integrated intensity from 532 to 540 eV.
    \label{Figure1}
    }
\end{figure}

\Tp{}-${\text{La}_\text{1.8-x}\text{Eu}_\text{0.2}\text{Sr}_\text{x}\text{CuO}_\text{4}}$ ($x$ = 0, 0.05, and 0.10) powders were synthesized through four distinct processes and densely sintered as described elsewhere~\cite{Takamatsu_Appl.Phys.Express57_2012}.
As-grown samples were non-SC and the $x$ = 0 sample was annealed and became SC below the \Tc{} of 20 K.
HAXPES measurements were performed at beamline 47XU of SPring-8 at $T$ = 300 K using \hv{} = 7.94 keV nearly grazing-incidence linearly polarized light with the total energy resolution of 0.3 eV.
XAS measurements were conducted in the total electron yield mode at beamline 07LSU of SPring-8 using linearly polarized light at $T$ = 300 K.

In Fig.~\ref{Figure1} (a), we compare the HAXPES spectra of the O 1\textit{s} core level on four LESCO samples with Sr content $x$ = 0 (annealed and as-grown), 0.05 (as-grown), and 0.10 (as-grown).
It has been proved in the case of annealed NCCO thin films that the changes in the core-level binding energies are largely contributed from the chemical-potential shift.
The effects of annealing and Sr substitution can be seen in Fig.~\ref{Figure1} (a), which indicates the core-level peak position of each spectrum.
One can clearly observe a systematic shift: by annealing (Sr substitution), the core-level peak moves towards higher (lower) binding energies maintaining the line shape, meaning an upward (downward) shift of the chemical potential due to electron (hole) doping.
Note that although there are signals of contamination located at higher binding energies because of sintered polycrystalline samples, these additional hump features are well separated from the main O 1$\textit{s}$ peaks and have no effects on the measured core-level shifts.
We could not use the spectra of La 3\textit{d} core level to deduce the chemical-potential shift because contamination signals overlap the main peaks.

Figure~\ref{Figure1} (b) shows the O $\textit{K}$-edge XAS spectra. 
Irrespective of annealing or Sr substitution, the spectra preserve their line shapes and do not show significant changes at the pre-edge peaks at 528 - 529 eV. 
If the CT gap exists, the pre-edge peak corresponds to transition into the upper Hubbard band in electron-doped cuprates~\cite{Pellegrin_Phys.Rev.B476_1993}, or transition into the ZRS that rapidly grows with increasing hole concentration in hole-doped cuprates~\cite{Chen_Phys.Rev.B8813_2013,Chen_Phys.Rev.Lett.661_1991,Pellegrin_Phys.Rev.B476_1993}. 
The latter possibility is ruled out since the pre-edge peak does not change irrespective of Sr substitution.
In the case of collapsed CT gap~\cite{Adachi_J.Phys.Soc.Jpn.826_2013}, the pre-edge peak represents the unoccupied part of the band above \EF{}.

\begin{figure}[!bt]
    \centering
    \includegraphics[width=0.5\textwidth]{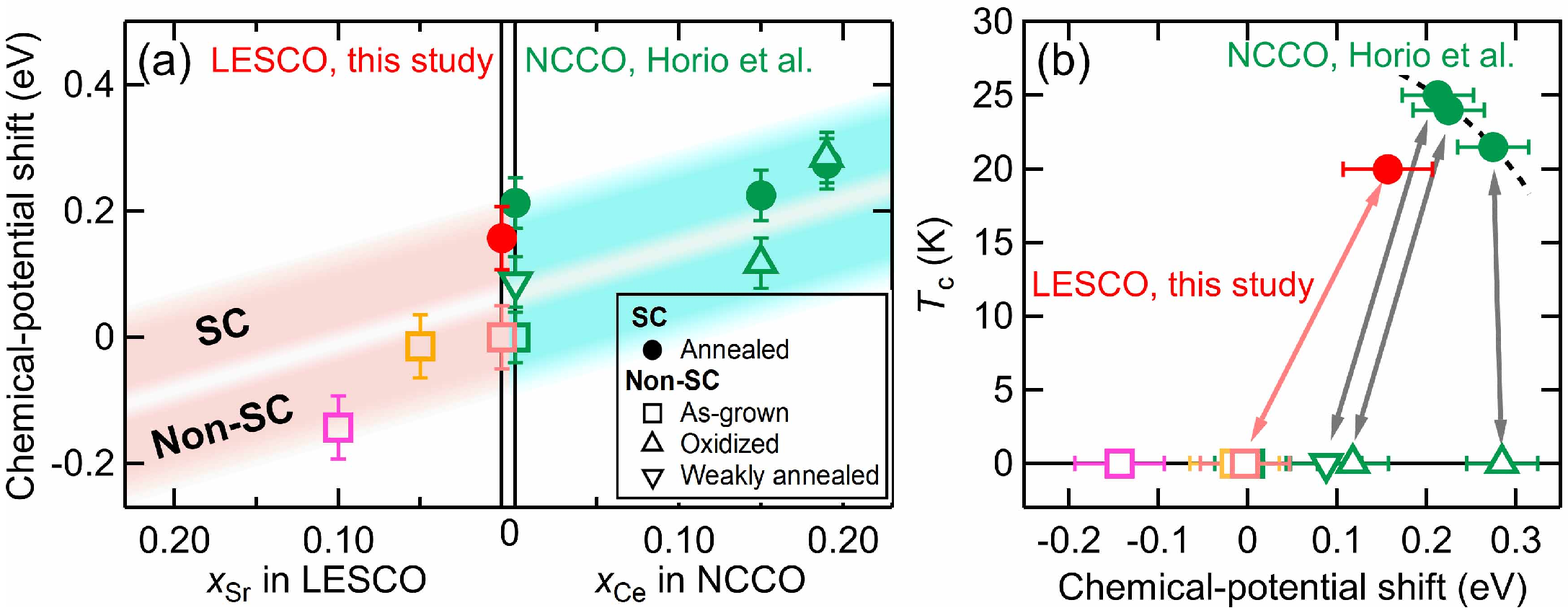}
    \caption{Chemical-potential shift in \Tp{}-${\text{La}_\text{1.8-x}\text{Eu}_\text{0.2}\text{Sr}_\text{x}\text{CuO}_\text{4}}$ (LESCO) bulk crystals and ${\text{Nd}_\text{2-x}\text{Ce}_\text{x}\text{CuO}_\text{4}}$ (NCCO) thin films~\cite{Horio_Phys.Rev.Lett.12025_2018}.
    (a) Chemical-potential shift defined as the O 1\textit{s} core-level shift plotted against Sr concentration $x_{\textup{Sr}}$ for LESCO (left) and chemical-potential shift in NCCO thin films plotted against Ce concentration $x_{\textup{Ce}}$ for comparison (right)~\cite{Horio_Phys.Rev.Lett.12025_2018}.
    Annealed samples (filled symbols) are superconducting (SC) while as-grown, weakly annealed, and oxidized samples (open symbols) are non-SC.
    (b) \Tc{} plotted against the chemical-potential shift for LESCO and NCCO~\cite{Horio_Phys.Rev.Lett.12025_2018}.
    Arrows connect SC and non-SC samples with the same Ce dopant levels.
    \label{Figure2}
    }
\end{figure}

To visualize the shift of chemical potential and compare them with other electron-doped cuprates, in Fig.~\ref{Figure2} (a), we plot the chemical-potential shift against dopant concentration for LESCO and NCCO~\cite{Horio_Phys.Rev.Lett.12025_2018}.
The shift of the chemical potential is evaluated from the core-level peak shift of the O 1\textit{s} spectra, taking the data of as-grown $x$ = 0 as the reference.
As shown in the figure, an upward (downward) chemical-potential shift of \wave{} 0.15 eV is realized by reduction annealing (by 10\% Sr substitution).
If a CT gap exists, when the system evolves from one type of carrier to another, one would expect a large chemical-potential jump, e.g., \wave{} 0.8 eV for ${\text{Y}_\text{0.38}\text{La}_\text{0.62}\text{Ba}_\text{1.74}\text{La}_\text{0.26}\text{Cu}_\text{3}\text{O}_\text{y}}$ ~\cite{Ikeda_Phys.Rev.B822_2010}.
Therefore there is no chemical-potential jump in the LESCO system within the doping range studied in the present work.
Meanwhile, in the presence of a CT gap, the absence of the ZRS in the XAS spectra suggests that there are no doped holes irrespective of Sr content.
Consequently, it is likely that the system remains on the electron-doped side of the Mott insulator.
Alternatively, if there is no CT gap~\cite{Adachi_J.Phys.Soc.Jpn.826_2013}, one can explain the present data, too 
The \Tc{} of LESCO and NCCO plotted against the chemical-potential shift in Fig.~\ref{Figure2} (b) further confirms the scenario that electron doping induced by reduction annealing is necessary to realize the superconductivity in non-Ce-doped \Tp{}-type cuprates~\cite{Horio_Phys.Rev.Lett.12025_2018,Horio_Phys.Rev.B982_2018}.

In conclusion, in \Tp{}-LESCO, electrons and holes are doped by reduction annealing and Sr substitution, respectively. 
In the presence of a CT gap, the system remains on the electron-doped side of the Mott insulator, while the present data can also be understood if the CT gap collapsed.

\begin{acknowledgments}
HAXPES and XAS experiments were performed at SPring-8 (Proposals Nos. 2016A1210 and 2018A1073).
This work was supported by KAKENHI Grants (Nos. 14J09200, 15H02109, 17H02915, and 19K03741) from JSPS.
\end{acknowledgments}

\bibliography{LESCO}

\end{document}